\documentclass{cseg2013}
\usepackage{xy}
\usepackage{amsmath,graphicx,natbib}
\usepackage{rotating}
\usepackage{lscape}
\usepackage{color}
\definecolor{tcolor}{rgb}{0.89,0.42,0.039} 
\usepackage{floatrow} 

\begin{document}
\begin{center}
\vspace*{-0.8in}
\includegraphics[height=2cm]{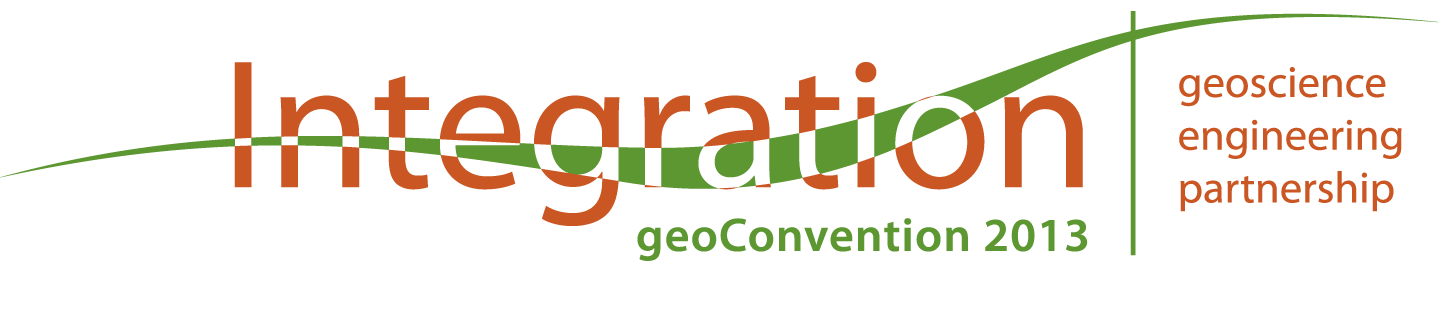}\\
\vspace{12pt}
 \Large \color{tcolor}{\bf Time-Frequency Representation of Microseismic Signals using the Synchrosqueezing Transform}\\
\vspace{10pt} \small \color{black} Roberto H. Herrera$^*$, Jean-Baptiste Tary and Mirko van der Baan. \\ University of Alberta, Edmonton, Canada. rhherrer@ualberta.ca
\vspace{23pt}
\end{center}

\section{Summary}
Resonance frequencies can provide useful information on the deformation occurring during fracturing experiments or $CO_2$ management, complementary to the microseismic event distribution. An accurate time-frequency representation is of crucial importance prior to interpreting the cause of resonance frequencies during microseismic experiments. The popular methods of Short-Time Fourier Transform (STFT) and wavelet analysis have limitations in representing close frequencies and dealing with fast varying instantaneous frequencies and this is often the nature of microseismic signals. The synchrosqueezing transform (SST) is a promising tool to track these resonant frequencies and provide a detailed time-frequency representation.
Here we apply the synchrosqueezing transform to microseismic signals and also show its potential to general seismic signal processing applications.

\section{Introduction}
Traditional time frequency representations, such as the Short-Time Fourier Transform (STFT) and the Wavelet Transform (WT) and special representations like Empirical Mode Decomposition (EMD), have limitations when signal components are not well separated in the time-frequency plane \citep{Wu2011}.  Synchrosqueezing, first introduced in the context of speech signals \citep{Daubechies1996} has shown to be an alternative to the EMD method \citep{Daubechies2011}, improving spectral resolution.

On the other hand, these improvements in time-frequency representation can be applied to microseismic signals to increase the readability of the frequency spectrum \citep{Auger1995} and identify independent components \citep{Thakur2012}.
An accurate representation of resonance frequencies could help gain a better understanding of the fracturing process, as it is challenging to determine what is exactly beneath the Earth's surface during the hydraulic fracturing process.

In this paper we demonstrate the potentiality of the SST in time frequency representation of microseismic signals. First we describe the underlying modulation model for microseismic signals and then we validate the method with synthetic examples. Finally we apply SST to a real data sample with high amplitude noise.
\section{Theory}
Coherent signals in microseismic recordings can have both anthropogenic and natural sources. In both cases the signal has similar oscillatory characteristics and could be represented as the sum of individual time-varying harmonic components:
\begin{equation}\label{eq:model}
 s(t)= \sum_{k=1}^K A_k(t) cos(\theta_k(t)) + \eta(t),
\end{equation}
\noindent where $A_k(t)$ is the instantaneous amplitude and $f_k(t)=\frac{1}{2\pi}\frac{d}{dt}\theta_k$ is the
instantaneous frequency (IF) of the resonator $k$. $\eta(t)$ represents the additive noise, including the contribution of environmental and acquisition sources. $K$ stands for the maximum number of components in one signal.

Thus, microseismic signals show some similarities with the time-frequency representation (TFR) of speech signals \citep{Daubechies1996}. The number of harmonics or components in the signal ($K$)
is random. They can appear at different time slot, with different amplitude ($A_k(t)$) and instantaneous frequency ($f_k(t)$).

\emph{From CWT to SST}\\
The CWT of a signal $s(t)$ is \citep{Daubechies1992}:
\begin{equation}\label{eq:CWT}
    W_s(a,b)= \frac{1}{\sqrt{a}}\int s(t)\psi^*(\frac{t-b}{a})dt,
\end{equation}
\noindent where $\psi^*$ is the complex conjugate of the mother wavelet, $b$ is the time shift applied to the mother wavelet which is also scaled by $a$. The CWT is simply the cross-correlation of the signal $s(t)$ with a number of wavelets that are scaled and translated versions of the original mother wavelet. $W_s(a,b)$ are the coefficients representing a concentrated time-frequency picture, which is used to extract the instantaneous frequencies \citep{Daubechies2011}.

The wavelet coefficients $W_s(a,b)$, often spreads out around the scale dimension $a$, leading to a blurred projection in time-scale representation. \citet{Daubechies1996} show that if the smear in the b-dimension can be neglected, then the instantaneous frequency $\omega_s(a,b)$ can be computed as the derivative of the wavelet transform at any point $(a,b)$ for which $W_s(a,b)\neq 0$:
\begin{equation}\label{eq:instF_ab}
     \omega_{s}(a,b)= \frac{-j}{W_s(a,b)}\frac{\partial W_s(a,b)}{\partial b}.
\end{equation}
The final step in the new time-frequency representation is to map the information from the time-scale plane to the
time-frequency plane. Every point $(b,a)$ is converted to $(b,\omega_{s}(a, b))$, this operation is called synchrosqueezing \citep{Daubechies2011}. Since $a$ and $b$ are discrete values we can have a scaling step $\Delta a_k = a_{k-1} - a_k$ for any $a_k$ where $W_s(a,b)$ is computed. Likewise, when mapping from the time-scale plane to the time-frequency plane $(b,a) \rightarrow (b,w_{inst}(a,b))$, the synchrosqueezing transform $T_s(w,b)$ is determined only at the centers $\omega_l$ of the frequency range $[\omega_l - \Delta \omega/2,\omega_l + \Delta \omega/2]$, with $\Delta \omega = \omega_l - \omega_{l-1}$:
\begin{equation}\label{eq:SST}
     T_{s}(\omega_l,b)= \frac{1}{\Delta \omega} \sum_{a_k:|\omega(a_k,b)- \omega_l| \leq \Delta \omega/2} W_s(a_k,b)a^{-3/2} \Delta a_k.
\end{equation}
The above equation shows that the time-frequency representation of the signal $s(t)$ is synchrosqueezed along the frequency (or scale) axis only \citep{Li2012}. The synchrosqueezing transform reallocates the coefficients of the continuous wavelet transform to get a concentrated image over the time-frequency plane, from which the instantaneous frequencies are then extracted \citep{Wu2011}. This is an ultimate goal in microseismic signal analysis. The identified frequencies are used to describe their source mechanisms and eventually gain a better understanding of the reservoir deformation.

\subsection*{\small Synthetic example}
\begin{figure}[htb]
\floatbox[{\capbeside\thisfloatsetup{capbesideposition={right,center},capbesidewidth=8cm}}]{figure}[\FBwidth]
{\caption{Synthetic Example. a) Synthetic signal with four sinusoidal components, b) the instantaneous frequencies obtained as the derivative of each one of the independent components. c) The STFT of the synthetic signal. A Hanning window of 321 samples and 50 \% overlap was used. Note the smearing effect in the bandwidth. In d) the SST shows a sharper representation of the instantaneous frequencies.}\label{fig1}}
{\includegraphics[width=8.5cm,height=9.5cm]{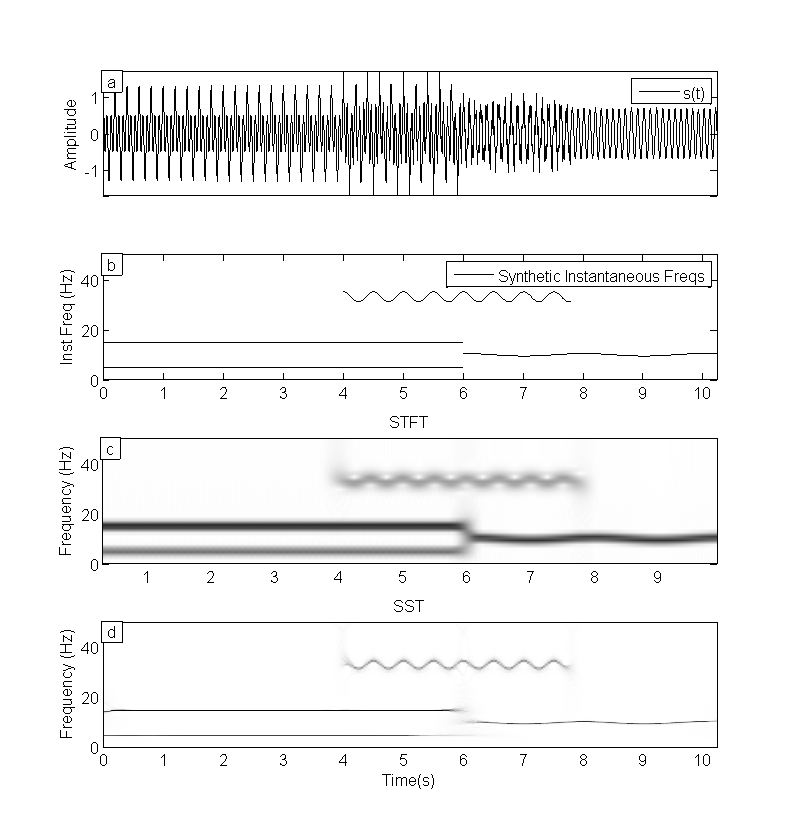}}
\end{figure}

Following the modulation model in Equation \ref{eq:model} we create a noiseless synthetic signal
as the sum of the following components (the IF is in brackets):
\begin{eqnarray*}
  s_1(t) &=& 0.3 \cos(10 \pi t); [ IF = 5 ]; \\
  s_2(t) &=& 0.8 \cos(30 \pi t); [ IF = 15 ]\\
  s_3(t) &=& 0.7 \cos(20 \pi t + \sin(\pi t)); [ IF = 10 + \cos(\pi t)/2 ]; \\
  s_4(t) &=& 0.4 \cos(66 \pi t + \sin(4 \pi t)); [ IF = 33 + 2 \cos(4\pi t) ]; \\
\end{eqnarray*}
The synthetic signal $s(t)$ (Figure \ref{fig1}a) has two constant harmonics of 5 Hz ($s_1(t)$) and 15 Hz ($s_2(t)$) from time 0 to 6 s, the 10 Hz component ($s_3(t)$) is modulated by a 0.5 Hz sinusoid from 6 to 10 s and a 2nd modulated signal with a central frequency of 33 Hz ($s_4(t)$) appears at time 4 s and vanishes at 7.8 s.

Figure \ref{fig1}b) shows the instantaneous frequencies for each component.
The STFT is able to identify the four components but with a low resolution (see Figure \ref{fig1}c), especially when they overlap at 6 s. On the other hand, the SST (Figure \ref{fig1}d) is able to perfectly delineate each individual component and to resolve the instantaneous frequencies close to the theoretical value, including $s_3(t)$ and $s_4(t)$ which are amplitude modulated.

\vspace*{-0.3in}
\section{\small Application to real microseismic signals}
In this section we apply the SST to a real dataset. A two-stage microseismic experiment was recorded by twelve 3C geophones deployed in a vertical well. In both stages the sampling frequency was 4000 Hz. The receivers and fluid injection are located approximately at the same depth.
The SST will be challenged by typical problems encountered in microseismic experiments analysis, i.e, time and frequency resolutions, noise stability and capability to identify localized frequency variations.

\begin{figure}[htb]
\floatbox[{\capbeside\thisfloatsetup{capbesideposition={right,center},capbesidewidth=8cm}}]{figure}[\FBwidth]
{\caption{Real example. Upper plot shows the STFT and the lower plot the SST. The SST is able to delineate the spectral components
some of them were missing in the STFT representation.}\label{fig2}}
{\includegraphics[width=8.5cm,height=8cm]{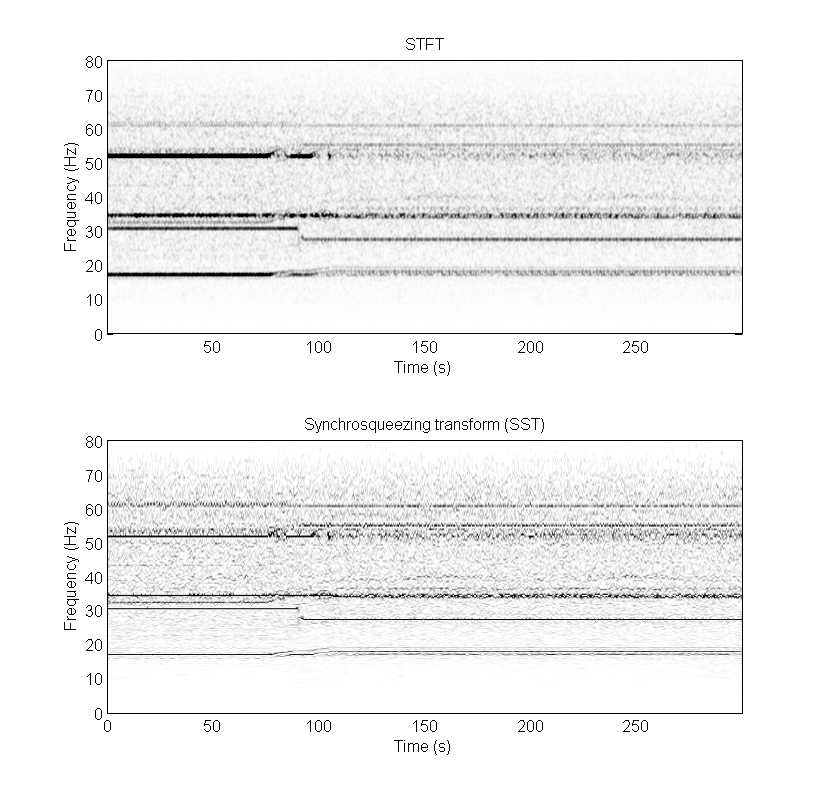}}
\end{figure}

A segment of 5 minutes of data, presenting clear resonance frequencies as well as sharp
and smooth changes in frequency content, was chosen to challenge the performance of the SST
representation. The STFT will be used as a reference, see Figure \ref{fig2} upper plot.
The STFT was calculated using a window length of 2 s with 50 \% overlap, involving a time resolution
of the same order. Four main resonance frequencies are clearly visible at approximately 18, 31, 35 and 52 Hz (Figure \ref{fig2} top).
Two additional lines of lower amplitude are also visible at 60 Hz and between the lines at 31 and 35 Hz.
At approximately 90 s, the amplitude of all lines becomes weaker, the one at 52 Hz almost vanishes. The line at
31 Hz abruptly drops down to 27 Hz while smooth changes are barely discernible on the 18 Hz line. The fuzzy background is not only
due to the high amplitude noise present in the data but also to spectral leakage introduced by the Fourier transform.

Despite the high-amplitude noise, the resonance frequencies as well as their variations are all clearly visible on the SST representation (Figure \ref{fig2} bottom).
The SST is able to map smooth and sharp changes in the frequency lines.
Compared with the STFT, the SST brings out the resonance frequencies more sharply,
improving significantly the frequency resolution.
The SST is then able to distinguish between the different lines constituting the resonance frequency at 18 Hz
while the STFT cannot. Last but not least, the SST determines the IFs at all times.
Therefore, the time resolution of this method is not limited by the size of any window.

For the SST we use a bump wavelet with a ratio central frequency to bandwidth of 50.
The discretization of the scales of CWT was 64. The SST produces a sharper time-frequency representation showing frequency
components that were hidden in the STFT representation.

\section{Conclusions}
We have introduced the Synchrosqueezing transform with application to microseismic signal analysis. This new transform shows promising results in the identification of resonant components and hence in the explanation of microseismic phenomena. The enhancement in spectral resolution allows the separation of close spectral bands while the short-time Fourier transform smears out these frequency components. SST is therefore attractive for high-resolution time-frequency analysis of microseismic signals.

\section{Acknowledgements}
The authors thank the sponsors of the BLind Identification of Seismic Signals (BLISS) Project and the Microseismic Industry Consortium for their financial support. The authors gratefully acknowledge an anonymous company for permission to show and use their data.

\bibliographystyle{seg} 
\bibliography{refsynchro}

\end{document}